\documentclass[aps,prb,groupedaddress,twocolumn]{revtex4}
\usepackage{color}
\usepackage{graphicx}
\usepackage{dcolumn}
\usepackage{bm}
\usepackage{ulem}

\begin{document}

\title{First-order chiral to non-chiral transition in the angular dependence of the upper critical induction of the Scharnberg-Klemm $p$-wave pair state}

\author{Jingchuan Zhang$^{1,2}$}
\email[]{nuscire@gmail.com}
\author{Christopher L{\"o}rscher$^{2}$}
\author{Qiang Gu$^{1}$}
\author{Richard A. Klemm$^{2}$}
\email[]{richard.klemm@ucf.edu}

\affiliation{$^1$Department of Physics, University of Science and Technology Beijing, Beijing 100083, China\\
$^2$Department of Physics, University of Central Florida,  Orlando, FL 32816-2385 USA}

\date{\today}

\begin{abstract}
We calculate the temperature $T$ and angular $(\theta,\phi)$ dependence  of the upper critical induction $B_{c2}(\theta,\phi,T)$ for parallel-spin superconductors with an axially symmetric $p$-wave pairing interaction pinned to the lattice and a dominant ellipsoidal Fermi surface (FS).   For all FS anisotropies, the chiral Scharnberg-Klemm  state $B_{c2}(\theta,\phi,T)$  exceeds that of the chiral Anderson-Brinkman-Morel state, and  exhibits a kink at $\theta=\theta^{*}(T,\phi)$, indicative of a first-order transition from its chiral, nodal-direction   behavior to its non-chiral, antinodal-direction behavior.  Applicability to Sr$_2$RuO$_4$, UCoGe, and topological superconductors such as Cu$_x$Bi$_2$Se$_3$ is discussed.
\end{abstract}

\maketitle
\section{Introduction}
Recently, there has been a great deal of interest in $p$-wave superconductivity\cite{Huy,deVisser,HH,Levy,Yelland,Aoki2,AokiFlouquet,SK1985,Lorscher,Mackenzie,Maeno,Deguchi,Machida08,Kittaka,Choi,Yonezawa,Suderow,Machida13,FuBerg,QiZhang,Kriener,Bay,Kirschenbaum,NLevy}.  The most likely candidate $p$-wave superconductors are the ferromagnetic superconductors UGe$_2$, UCoGe, and URhGe, which exhibit long-range ferromagnetism well above the superconducting transition temperature $T_c$, and the same electrons participate in the ferromagnetism and the superconductivity\cite{Huy,deVisser,HH,Levy,Yelland,Aoki2,AokiFlouquet}.  In URhGe, measurements of the temperature $T$ dependence of the upper critical induction $B_{c2}(T)$ in the three crystal axis directions was found to fit the Scharnberg-Klemm  theory of the $p$-wave polar state with completely broken symmetry (CBS)\cite{HH,SK1985}, with single-component $p_z$-pairing state only along the crystal $a$-axis. Subsequent experiments found a reentrant superconducting phase at much higher magnetic field ${\bm H}$ strengths,  violating the conventional Pauli limit $B_P=1.85T_c$ (T/K) by  a factor of 20.  $B_{c2}$ in UCoGe also violates $B_P$ by a factor of 20, but its anisotropy suggests that if the superconductivity were $p$-wave, it would be more likely to have an axial state form, such as do the chiral Anderson-Brinkman-Morel (ABM) and chiral Scharnberg-Klemm (SK) states\cite{AM,AB,SK1980,Mineev}. Second, there has been an even greater interest in Sr$_2$RuO$_4$, as the Knight shift measurements for ${\bm H}$ parallel and perpendicular to the layers all showed no temperature $T$ dependence below  $T_c$, suggestive of a parallel-spin state\cite{Mackenzie,Maeno}.  However, $B_{c2}$ experiments on that material were shown to be strongly Pauli limited for ${\bm B}\perp\hat{\bm c}$\cite{Deguchi,Machida08,Kittaka,Choi,Yonezawa,Machida13,Zhang}, and scanning tunneling microscopy experiments showed strong evidence for a nodeless gap\cite{Suderow}, although with cylindrical Fermi surfaces (FSs), this might be consistent with an axial $p$-wave state.  Third, there has been a large recent interest in topological insulators, in the hope that they might become chiral $p$-wave superconductors with doping, applied pressure, or proximity coupling\cite{FuBerg,QiZhang,Kriener,Bay,Kirschenbaum,NLevy}. Initial $B_{c2}(T)$ measurements on Cu$_x$Bi$_2$Se$_3$ were consistent with a $p$-wave polar state for ${\bm H}$ both parallel and perpendicular to the layers\cite{Bay,SK1980}. However, scanning tunneling microscopy (STM) experiments established that Cu$_x$Bi$_2$Se$_3$ has an isotropic gap strongly suggestive of an $s$-wave order parameter (OP)\cite{NLevy}, and that isotropic $s$-wave OP was respectively proximity-induced up to 7 K and 50 K into   Bi$_2$Se$_3$ layers deposited atop the $c$-axes of the layered low-$T_c$ and high-$T_c$ superconductors, 2$H$-NbSe$_2$ and Bi$_2$Sr$_2$CaCu$_2$O$_{8+\delta}$\cite{Shanghai,Wang}, consistent with $s$-wave substrate crystal OPs in the $c$-axis direction of both of those layered superconductors\cite{book,Mueller,Li,PMReview}. However,  still undiscovered topological superconductors  might have axial $p$-wave OP symmetry.

Previously, we generalized the microscopic calculation of $B_{c2}(T)$ for the $p$-wave polar state pinned to a crystal lattice direction to  extend its validity to a superconductor with a dominant ellipsoidal FS and ${\bm B}$ in an arbitrary direction, ${\bm B}=B(\sin\theta\cos\phi,\sin\theta\sin\phi,\cos\theta)$, with respect to the crystal lattice, in order to provide a sound theoretical basis for a more sensitive probe of the actual OP in orthorhombic materials such as URhGe.  Here we use the same technique to construct a theory of the full angular dependence of $B_{c2}(\theta,\phi,T)$ for the ABM and SK states, in order to identify the symmetry of the OP in  UCoGe, Sr$_2$RuO$_4$, and other candidate materials.  Since UCoGe is orthorhombic, the ellipsoidal FS model is the best that can be made without additional features such as magnetic pairing fluctuation effects and ${\bm B}$ dependencies of the pairing interactions\cite{HattoriT}, or ${\bm B}$-dependent interactions\cite{HattoriK}, {\it etc.}  For tetragonal Sr$_2$RuO$_4$, the lack of any detectable ferromagnetism strongly suggests weak coupling interactions, but there are three barrel shaped FSs, and the STM experiments strongly suggest nearly equal isotropic gaps on each\cite{Suderow}. Although one could envision a scenario in which one FS dominated $B_{c2}(0^{\circ},\phi,T)$ and another dominated $B_{c2}(90^{\circ},\phi,T)$, since the latter is of primary interest,  it suffices to consider only one FS. Moreover, as the $k_z$ dispersion of those bands is sufficient to avoid dimensional crossover effects in $B_{c2}(90^{\circ},\phi,T)$ measurements\cite{Kittaka,book,KBL,KS}, an ellipsoid of uniaxial anisotropy is sufficient to examine $B_{c2}$ measurements for all ${\bm B}$ directions with high accuracy\cite{book}.  As anticipated earlier, for a parallel-spin pairing interaction of the form $V(\hat{\bm k},\hat{\bm k}')=3V_0(\hat{k}_1\hat{k}_1'+\hat{k}_2\hat{k}_2')$, one would expect $B_{c2}(\theta,\phi,T)$ to be given by the SK state\cite{SK1980,Lorscher}. Although a favorite pair state for Sr$_2$RuO$_4$ has the form $\hat{\bm z}(\hat{k}_1+i\hat{k}_2)$, where the ${\bm d}$-vector $\hat{\bm z}$ corresponds to the antiparallel-spin state in the lattice representation, we shall here assume that the spins are parallel\cite{SK1980}, and will include Pauli limiting effects subsequently\cite{Zhang}.  Here we present detailed calculations of the $B_{c2}(\theta,\phi,T)$ for both the ABM and SK states on a single ellipsoidal FS.

We  assume weak
coupling for a clean homogeneous type-II parallel-spin $p$-wave superconductor with  effective Hamiltonian  $~$\cite{SK1980,Lorscher},
\begin{eqnarray}
\cal{H}&=&\sum_{{\bm k},\sigma=\pm}a_{{\bm k},\sigma}^{\dag}[\epsilon({\bm k}-e{\bm A})-\mu]a^{}_{{\bm k},\sigma}\nonumber\\
&&+\frac{1}{2}\sum_{{\bm k},{\bm k}',\sigma}a_{{\bm k}',\sigma}^{\dag}a_{{\bm k},\sigma}^{\dag}V(\hat{\bm k},\hat{\bm k}')a^{}_{{\bm k},{\sigma}}a^{}_{{\bm k}',{\sigma}},\\
V(\hat{\bm k},\hat{\bm k}')&=&\frac{3}{2}V_{0}\sum_{\sigma'=\pm}f_{\sigma'}(\hat{\bm k})\hat{\bm d}_{\sigma'}\cdot\hat{\bm d}_{\sigma'}^{*}f^{*}_{\sigma'}(\hat{\bm k}'),
\label{interaction}
\end{eqnarray}
where we assume parallel-spin pairing with $\hat{\bm d}_{\sigma'}=\hat{\bm x}+i\sigma'\hat{\bm y}$ and  $f_{\sigma'}(\hat{\bm k})=(\hat{k}_1+i\sigma'\hat{k}_2)$ from the degenerate $\Gamma_3^{-}$ and $\Gamma_4^{-}$ tetragonal point group representations\cite{Sigrist}, $e$ is the electronic charge,  $\mu$ is the chemical potential, the unit wave vectors $\hat{k}_i$ were previously defined on an ellipsoidal FS\cite{Lorscher},
and we set $\hbar=k_B=1$.   For non-ferromagnetic candidate $p$-wave superconductors, the upper critical induction $B_{c2}=\mu_0H_{c2}$, where $H_{c2}$ is the upper critical field.
After performing the Klemm-Clem (KC) transformations\cite{KC} that map the ellipsoidal FS onto a spherical one and then rotate the transformed induction to the new $\tilde{z}$ axis direction, the transformed linear gap equation becomes
\begin{eqnarray}
\overline{\tilde{\Delta}}(\tilde{\bm R},\hat{\tilde{\bm k}})&=& T\sum_{\omega_{n}}\frac{N(0)}{2}\int d\Omega_{\tilde{\bm k}'}\tilde{V}(\hat{\tilde{\bm k}},\hat{\tilde{\bm k}'})\int_0^{\infty}d\xi_{\tilde{\bm k}'}\nonumber\\
& &\times e^{-2\xi_{\tilde{\bm k}'}|\omega_{n}|}e^{-i\xi_{\tilde{\bm k}'}v_F\hat{\tilde{\bm k}'}\cdot\tilde{\bm\Pi}(\tilde{\bm R})}\overline{\tilde{\Delta}}(\tilde{\bm R},\hat{\tilde{\bm k}'}),\label{gapequation}
\end{eqnarray}
where $\overline{\tilde{\Delta}}$ is the transformed $\Delta$ amplitude without the gauge phases\cite{Lorscher}, $N(0)=mk_F/(2\pi^2)$ is the density of states per spin at the chemical potential $\mu$ for an effectively isotropic metal with a geometric mean mass $m=(m_1m_2m_3)^{1/3}$, effective Fermi wave vector $k_F=\sqrt{2m\mu}$,  effective  Fermi velocity $v_F=k_F/m$, and
$\tilde{\bm\Pi}(\tilde{\bm R})=-i\alpha\tilde{\bm\nabla}_{\tilde{\bm R}}-2e\tilde{\bm A}(\tilde{\bm R})$,
where $\alpha(\theta,\phi)=\sqrt{\overline{m}_3}\sqrt{\cos^2\theta+\gamma^{-2}(\phi)\sin^2\theta}$, $\overline{m}_i=m_i/m$, and
 $\gamma^2(\phi)=\frac{m_3}{m_1\cos^2\phi+m_2\sin^2\phi}$ is the ellipsoidal anisotropy function\cite{Lorscher}.
The KC transformations change $V(\hat{\bm k},\hat{\bm k}')$ in Eq. (\ref{interaction}) to
\begin{eqnarray}
\tilde{V}(\hat{\tilde{\bm k}},\hat{\tilde{\bm k}'})&=&\frac{3}{2}V_0
\sum_{\sigma=\pm}\tilde{f}_{\sigma}(\hat{\tilde{\bm k}})\tilde{f}^{*}_{\sigma}(\hat{\tilde{\bm k}}')\label{Vtransformed}
\end{eqnarray}
 where $\tilde{f}_{\sigma}(\hat{\tilde{\bm k}})=\hat{\tilde{k}}_1+i\sigma(\hat{\tilde{k}}_2\cos\theta'+\hat{\tilde{k}}_3\sin\theta')$, $\cos\theta'=\sqrt{\overline{m}_3}\cos\theta/\alpha$, {\it etc.}\cite{Lorscher}
From the form of $\tilde{V}(\hat{\tilde{\bm k}},\hat{\tilde{\bm k}}')$,  $\overline{\tilde{\Delta}}(\tilde{\bm R},\hat{\tilde{\bm k}})=\sum_{\sigma=\pm}\overline{\tilde{\Delta}}_{\sigma}(\tilde{\bm R})\tilde{f}_{\sigma}(\hat{\tilde{\bm k}})$,  we expand the $\overline{\tilde{\Delta}}_{\sigma}(\tilde{\bm R})$ in terms of the harmonic oscillator eigenfunctions $|n(\tilde{\bm R})\rangle$, $\overline{\tilde{\Delta}}_{\sigma}(\tilde{\bm R})=\sum_{n=0}^{\infty}a_n^{\sigma}|n(\tilde{\bm R})\rangle$, perform the integrals over the $\hat{\tilde{k}}_i'$ variables in the linearized gap equation, and obtain this double recursion relation for the $a^{(\pm)}_n$,
\begin{eqnarray}
 a_{n}^{(\pm)}&=&\Bigl(\frac{1}{2}(1+\cos^2\theta')a_{n}^{(\pm)}+\frac{1}{2}\sin^2\theta'a_{n}^{(\mp)}\Bigr)\alpha_{n}^{(a)}\nonumber\\
 &&+\frac{1}{2}\sin^2\theta'\Bigl(a_{n}^{(\pm)}-a_{n}^{(\mp)}\Bigr)\alpha_{n}^{(p)}\nonumber\\
 &&+\Bigl(\frac{1}{4}\sin^2\theta'a_{n+2}^{(\pm)}+\frac{1}{4}(1\pm\cos\theta')^2a_{n+2}^{(\mp)}\Bigr)\beta_{n}\nonumber\\
 & &+\Bigl(\frac{1}{4}\sin^2\theta'a_{n-2}^{(\pm)}+\frac{1}{4}(1\mp\cos\theta')^2a_{n-2}^{(\mp)}\Bigr)\beta_{n-2},\label{recursion}
 \end{eqnarray}
 where
 \begin{eqnarray}
 \alpha_{n}^{(p,a)}&=&\pi T{\sum_{\omega_{n}}}\int_{0}^{\pi}d\theta_{\tilde{\bm k}'}\sin\theta_{\tilde{\bm k}'}\left(
3\cos^{2}\theta_{\tilde{\bm k}'},
\frac{3}{2}\sin^{2}\theta_{\tilde{\bm k}'}
\right)\nonumber\\
&&\times\int_{0}^{\infty}d\xi_{\tilde{\bm k}'} e^{-2\xi_{\tilde{\bm k}'}|\omega_{n}|}e^{-\eta_{\tilde{\bm k}'}/2}L_{n}(\eta_{\tilde{\bm k}'}),\\
\beta_{n}&=&\pi T\sum_{\omega_{n}}\int_{0}^{\pi}d\theta_{\tilde{\bm k}'}\frac{3}{2}\mathrm{sin^{3}}\theta_{\tilde{\bm k}'}\int_{0}^{\infty}d\xi_{\tilde{\bm k}'} e^{-2\xi_{\tilde{\bm k}'}|\omega_{n}|}\nonumber\\
&&\times e^{-\eta_{\tilde{\bm k}'}/2}(-\eta_{\tilde{\bm k}'})L_n^{(2)}(\eta_{\tilde{\bm k}'})[(n+1)(n+2)]^{-1/2},
\end{eqnarray}
where
\begin{eqnarray}
\eta_{\tilde{\bm k}'}&=&eB\alpha(\theta,\phi)v_{F}^{2}\xi^{2}_{\tilde{\bm k}'}\sin^{2}\theta_{\tilde{\bm k}'},
 \end{eqnarray}
$t=T/T_{c}$, $T_{c}=(2e^C\omega_{0}/\pi)\exp\left(-1/N(0)V_{0}\right)$, $\omega_{0}$ is a characteristic pairing cutoff frequency, $C\approx0.5772$ is  Euler's constant,  and
$L_{n}(z)$ and $L_n^{(2)}(z)$ are  a Laguerre and an associated Laguerre polynomial, respectively\cite{SK1980,Lorscher}.

For the chiral ABM state, the decoupled $a_n^{(\pm)}$ each satisfy
$a_n^{(\pm)}D_n=\Gamma_na_{n+2}^{(\pm)}+\Gamma_{n-2}a^{(\pm)}_{n-2}$, where $D_n=1-\frac{1}{2}(1+\cos^2\theta')\alpha_n^{(a)}-\frac{1}{2}\sin^2\theta'\alpha_n^{(p)}$ and $\Gamma_n=\frac{1}{4}\sin^2\theta'\beta_n$.  Solving this recursion relation, we obtain the continued fraction expression from which $B_{c2}(\theta,\phi,t)$ for the ABM state is  obtained numerically,
\begin{eqnarray}
D_0-\frac{\Gamma_0^2}{D_2-\frac{\Gamma_2^2}{D_4-\ldots}}&=&0.\label{ABM}
\end{eqnarray}
As for the polar/CBS state\cite{Lorscher}, one iteration is accurate to a few percent, but four or five iterations are needed for the accuracy necessary to observe the interesting effects.

\begin{figure}
\includegraphics[width=0.45\textwidth]{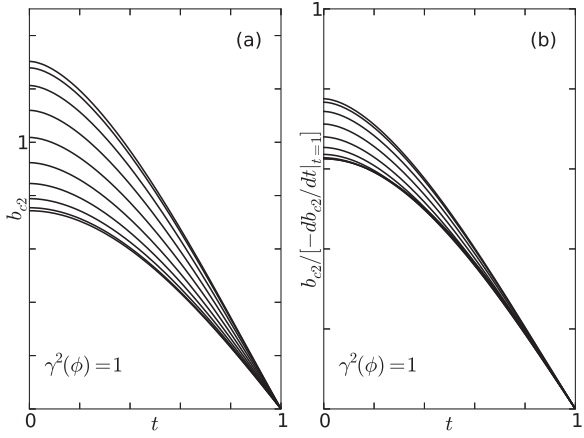}
\caption{(a) Reduced  $b_{c2}$ versus $t=T/T_c$ for the chiral ABM state [Eq. (\ref{ABM})] at $\theta$ values from $0^{\circ}$ (${\bm H}||\hat{\bm c}$, bottom) to $90^{\circ}$ (${\bm H}\perp\hat{\bm c}$, top), in increments of $10^{\circ}$ for a spherical FS. (b) Same curves normalized to have the same slopes at $T_c$.} \label{fig1}
\end{figure}

\begin{figure}
\includegraphics[width=0.45\textwidth]{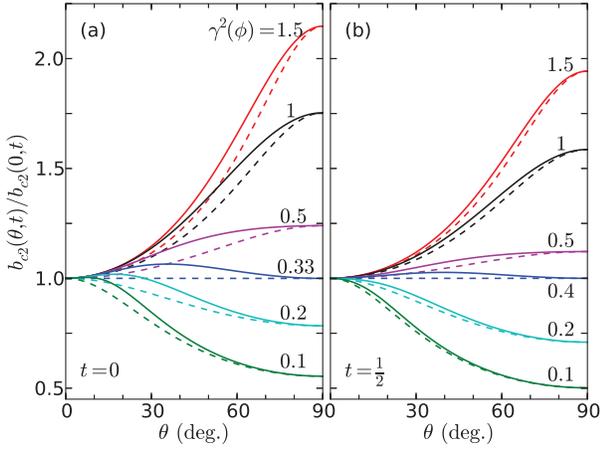}
\caption{(color online) Reduced  $b_{c2}$ versus $\theta$ for the chiral ABM state [Eq. (\ref{ABM})] at the indicated effective mass anisotropy $\gamma^2(\phi)$ values (solid) and the effective mass angular fits [Eq. (\ref{beff}), dashed] at $t=0$ (a) and $t=1/2$ (b).} \label{fig2}
\end{figure}

The results for the reduced $b_{c2}(\theta,t)$ for a parallel-spin superconductor in the $p$-wave ABM state with a dominant spherical $\gamma^2(\phi)=1$ FS are shown in Fig. 1.  In Fig. 1(a), the curves for $\theta=0^{\circ} ({\bm b}||\hat{\bm c})$ (nodal direction) to $90^{\circ} ({\bm b}\perp\hat{\bm c})$ (antinodal direction) are shown in increments of $10^{\circ}$.  The result for the nodal direction ($\theta=0^{\circ}$ was obtained previously\cite{SK1980}.  Just below $T_c$, $b_{c2}(\theta,\phi,t)\propto[m_3\cos^2\theta+2\gamma^{-2}(\phi)\sin^2\theta]^{-1/2}$, where the factor 2 arises from the ABM order parameter (OP) anisotropy.  In order to distinguish which part of the overall  $b_{c2}(\theta,t)$ anisotropy that is attributable solely to the order parameter anisotropy, in Fig. 1(b), those Fig. 1(a) results scaled to have the same slope at $t=1$ are presented.  Nothing unusual is evident from these spherical FS curves, and they are smooth and increase monotonically with increasing $\theta$.

However, we also studied the role of ellipsoidal (or uniaxial) FS anisotropy.  In Fig. 2, we chose fixed FS anisotropy values $\gamma^2(\phi)$ ranging from 0.1 to 1.5 and plotted in Figs. 1 (a,b) at $t=0$ and $\frac{1}{2}$, respectively.  The solid curves are evaluated from Eq. (\ref{ABM}). The dashed curves are the conventional ``effective mass'' anisotropy $b_{\rm eff}(\theta,t)$ forms obtained by fitting the calculated $b_{c2}(0^{\circ},t)$ and $b_{c2}(90^{\circ},t)$,
\begin{eqnarray}
b_{\rm eff}(\theta,t)&=&[\cos^2\theta/b^2_{c2}(0^{\circ},t)+\sin^2\theta/b^2_{c2}(90^{\circ},t)]^{-1/2}.\>\>\>\>\>\>\>\label{beff}
\end{eqnarray}
We note that $b_{c2}(\theta,t)$ exhibits  an unusual $\theta$ dependence, with a peak in  at $\theta^{*}$ for $\gamma^2(\phi)<\frac{1}{2}$ that is distinctly different than the conventional $b_{c2}$ maxima at $\theta=0^{\circ}$ or $90^{\circ}$.  Such anomalous double peaks at unconventional $\theta$ values satisfying $0<\theta^{*}<90^{\circ}$, and by reflection symmetry about $90^{\circ}$, also for $90^{\circ}<\theta^{*}<180^{\circ}$, were predicted earlier for the polar state pinned to the lattice\cite{Lorscher}.  However, in that case, the anomalous double peaks were predicted to occur for $\lambda(t)>\gamma^2(\phi)>3$, with maximal $\lambda(t)$ values for finite $t$.  Since the anomalous behavior is unlikely to be relevant to either Sr$_2$RuO$_4$ or UCoGe, for which  $\gamma^2\gg1$, for brevity, the $\lambda'(t)$ curve defining the lower limit of the range of $\theta^{*}$ for $\lambda'(t)<\gamma^2(\phi)<\frac{1}{2}$ will be presented elsewhere\cite{Zhang}.\\
\begin{figure}
\includegraphics[width=0.45\textwidth]{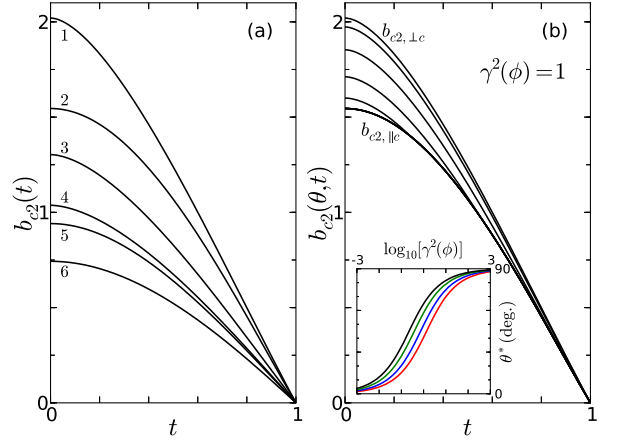}
\caption{(color online) (a) $b_{c2}(t)$ for  the antinodal SK state (1), nodal SK state (2), antinodal ABM state (3), $s$-wave state absent of Pauli limiting (4), planar nodal (CBS) state (5), and   nodal ABM state (6) on a spherical FS.  (b) Reduced  $b_{c2}(t)$  for the chiral SK state at $\theta$ values from $0^{\circ}$ (${\bm B}||\hat{\bm c}$, bottom) to $90^{\circ}$ (${\bm B}\perp\hat{\bm c}$, top), in increments of $10^{\circ}$ for a spherical FS. The $\theta=0^{\circ}$, $10^{\circ}$, $20^{\circ}$, $30^{\circ}$ and $40^{\circ}$ are indistinguishable on this scale.  Inset:  Plots of the kink angle $\theta^{*}$ versus $\log_{10}[\gamma^2(\phi)]$ from top to bottom for $t=\frac{3}{4}$ (black), $\frac{1}{2}$ (green), $\frac{1}{4}$ (blue), 0 (red). } \label{fig3}
\end{figure}
\begin{figure}
\includegraphics[width=0.45\textwidth]{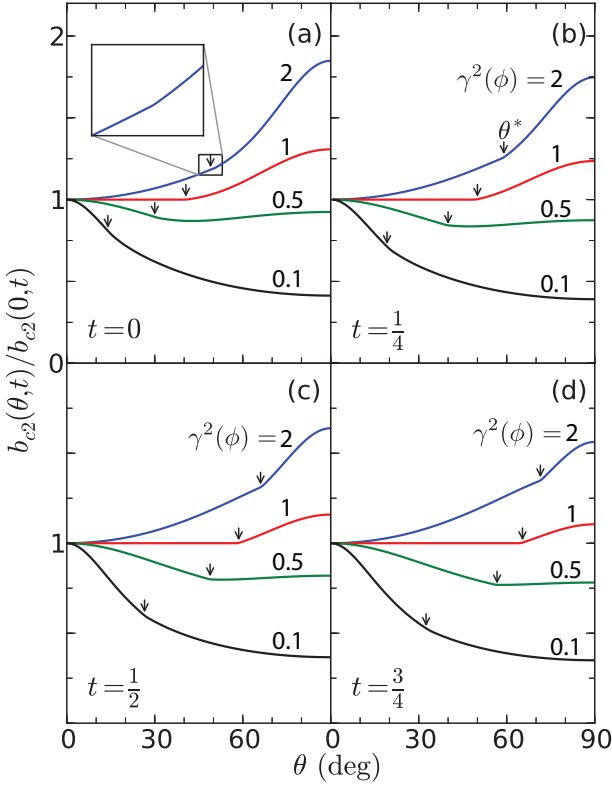}
\caption{(color online) Reduced upper critical induction $b_{c2}$ versus $\theta$ for the chiral SK state  for $\gamma^2(\phi)=2$ (blue, top), 1 (red), 0.5 (green), and 0.1 (black) at $t=0$ (a), $\frac{1}{4}$ (b), $\frac{1}{2}$ (c), and $\frac{3}{4}$ (d).  The arrows indicate kinks in $b_{c2}(\theta)$ at $\theta^{*}$, signifying first-order transitions from the chiral SK state ($\theta<\theta^{*}$) to the non-chiral antinodal SK (or polar) state $b_{c2}(t)$ curve ($\theta>\theta^{*}$).} \label{fig4}
\end{figure}
The much more interesting chiral axial $p$-wave state is the SK state.  We note that it is chiral as long as $\Delta^{(+)}\ne\Delta^{(-)}$, or  $a_n^{(+)}\ne a_n^{(-)}$ for at least one relevant $n$ value\cite{Lorscher}.  It is easy to see that for $\theta'=0$, Eq. {\ref{recursion}) reduces for $a_n^{(+)}\ne0$ to $[1-\alpha_n^{(a)}][1-\alpha_{n+2}^{(a)}]=\beta_n^2$, which for $a_n^{(+)}\ne0$ is the expression for the SK state with ${\bm B}$ in the nodal direction\cite{SK1980}, whereas for $\theta'=\pi/2$, it reduces for $a_n^{(+)}\ne a_n^{(-)}$ to $\alpha_n^{(p)}=1$, the expression for  the SK state with ${\bm B}$ in the antinodal (polar state) direction\cite{SK1980}.

However, for a general $\theta'$,  $a_n^{(+)}\ne a_n^{(-)}$, Eq. (\ref{recursion}) is a double recursion relation in the six harmonic oscillator amplitudes, $a_n^{(\pm)}$, $a_{n+1}^{(\pm)}$ and $a_{n-2}^{(\pm)}$, which requires further analysis to write the exact solution.  We first write $\Psi_n^{(\pm)}=\frac{1}{2}(a_n^{(+)}\pm a_n^{(-)})$, $D_n^{(+)}=1-\alpha_n^{(a)}$,  $D_n^{(-)}=1-\alpha_n^{(a)}\cos^2\theta'-\alpha_n^{(p)}\sin^2\theta'$,
 and construct $\phi_n^{(\pm)}=\cos\theta'D_n^{(+)}\Psi_n^{(+)}\pm D_n^{(-)}\Psi_n^{(-)}$. After letting $n\rightarrow n+2$ in the expression for $\phi^{(-)}_n$, we obtain
two equations for $\Psi_n^{(+)}$ and $\Psi_{n+2}^{(+)}$ in terms of $\Psi_n^{(-)}$ and $\Psi_{n+2}^{(-)}$. Using these equations to eliminate $\Psi_n^{(+)}$ and $\Psi_{n+2}^{(+)}$ in favor of $\Psi_n^{(-)}$ and $\Psi_{n+2}^{(-)}$,  letting $n\rightarrow n-2$ in the expression for $\Psi_{n+2}^{(-)}$, and equating that with the other expression for $\Psi_n^{(-)}$, we obtain the simple recursion relation for the $\Psi_n^{(-)}$, $A_n\Psi_{n+2}^{(-)}+B_n\Psi_n^{(-)}+C_n\Psi_{n+2}^{(-)}=0$, the solution of which may be expressed in the continued fraction equation,
\begin{eqnarray}
B_0-\frac{A_0C_0}{B_2-\frac{A_2C_2}{B_4-\ldots}}&=&0,\label{SK}
\end{eqnarray}
where $B_n=B_n^{(+)}-B_n^{(-)}$, $A_n=E_{n-2}\beta_n[\cos^2\theta'D_{n+2}^{(+)}-D_{n+2}^{(-)}]$,  $B_n^{(+)}=D_n^{(-)}[E_nD_{n-2}^{(+)}+E_{n-2}D^{(+)}_{n+2}]$, $B_n^{(-)}=\cos^2\theta'[\beta^2_nE_{n-2}+\beta^2_{n-2}E_n]$, $C_n=\beta_{n-2}E_n[\beta_{n-2}\cos^2\theta'D_{n-2}^{(+)}-D_{n-2}^{(-)}]$, and $E_n=D_n^{(+)}D_{n+2}^{(+)}-\beta_n^2$.  As for the polar/CBS state and the ABM state, one iteration is accurate to a few percent, but four or five iterations are necessary to display the most important features of this work.  We also eliminated $\Psi_n^{(-)}$ and $\Psi_{n+2}^{(-)}$ in favor of $\Psi_n^{(+)}$ and $\Psi_{n+2}^{(+)}$, but the $b_{c2}(\theta,\phi,t)$ values calculated from the resulting continued fraction equation  were always lower than those calculated from Eq. (\ref{SK}).

In Fig. 3(a), we plotted the reduced $b_{c2}(t)$ for the nodal and antinodal directions of the ABM, SK, and polar/CBS states, along with that [curve (4)] of a conventional $s$-wave superconductor without any Pauli limiting effects, all for a spherical FS.  The antinodal directions of the polar state and SK states both have $b_{c2}(t)$ curves described by curve (1), and the nodal direction of the SK state $b_{c2}(t)$ follows curve (2), as found previously\cite{SK1980}.  Curve (3) is the new $b_{c2}(t)$ curve for the antinodal direction of the ABM state.  Curves (5) and (6) describe the planar nodal polar/CBS state direction and the nodal direction of the ABM state, as also found previously\cite{SK1985}.  We note that the SK state $b_{c2}(\theta,\phi,t)$ is larger for all field directions than is the ABM state $b_{c2}(\theta,\phi,t)$, as the second chiral component of the OP allows for the state to be superconducting at larger applied field strengths.  In Fig. 3(b), the $t$ dependence of $b_{c2}(\theta,t)$ is illustrated for $\theta=0^{\circ}({\bm b}||\hat{\bm c})$ (bottom) to $\theta=90^{\circ} ({\bm b}\perp\hat{\bm c})$ (top), in increments of $10^{\circ}$.  Surprisingly, the curves for $\theta=0^{\circ}, 10^{\circ}, 20^{\circ}, 30^{\circ}$ and $40^{\circ}$ are remarkably close to one another, and appear to cross at finite $t$ values!  This is an indication of a chiral to non-chiral transition for $\theta\ge40^{\circ}$ at various $t$ values, as the vortices just below $b_{c2}$ appear to lock onto the nodal direction for $\theta\le40^{\circ}$, but for $\theta>40^{\circ}$ unlock from that direction, and favor the non-chiral antinodal (or polar) state direction.  Similar behavior was predicted recently for the  vortex structure in the mixed state of a chiral ABM state model of Sr$_2$RuO$_4$\cite{Machida13}.

To investigate this surprising feature in more detail, in Fig. 4 we show the $\theta$ dependence of $b_{c2}(\theta,\phi,t)$ at the effective mass anisotropy values $\gamma^2(\phi)=0.1, 0.5, 1,$ and 2, at $t=0, \frac{1}{4}, \frac{1}{2}$, and $\frac{3}{4}$.  In every case, there is a kink in $b_{c2}(\theta)$ at $\theta=\theta^{*}$ for fixed $\phi$ and $t$, which we interpret as evidence for  a first-order phase transition from a chiral to non-chiral state.  Although these kinks are easiest to see for small $\gamma^2$ values, and Sr$_2$RuO$_4$ has $\gamma^2>10^3$, our high-accuracy solutions of Eq. (\ref{SK}) allow us to determine $\theta^{*}[\gamma^2(\phi),t]$ with great precision.  In the inset to Fig. 3(b), we plotted $\theta^{*}$ in degrees versus $\ln_{10}[\gamma^2(\phi)]$ from -3 to 3 at the reduced $t$ values 0, $\frac{1}{2}, \frac{1}{2}$, and $\frac{3}{4}$.  Thus, if Sr$_2$RuO$_4$ were a chiral $p$-wave parallel-spin superconductor as often purported, then one ought to observe a first order chiral to non-chiral transition for $\theta\approx90^{\circ}$, nearly parallel to the layers.  It is therefore quite interesting to note that some evidence for this sort of behavior may have already been observed in very recent $H_{c2}(T)$ measurement on Sr$_2$RuO$_4$ \cite{Yonezawa}.  However, a cautionary note that is that $b_{c2}(90^{\circ},\phi,t)$ appears to be strongly Pauli limited\cite{Kittaka,Machida08,Deguchi,Choi}, and more details of such and other fits using this FS model will soon become available\cite{Zhang}.

With regards to the ferromagnetic superconductor UCoGe, the ferromagnetism in the $c$-axis direction allows for an axial-type parallel-spin $p$-wave pairing interaction, most likely mediated by ferromagnetic exchange interactions, in the $ab$ plane.  However, at large applied fields along the $b$-axis direction, not only does $B_{c2,b}(0)$ exceed the Pauli limit by a factor of at least 20, but the very strange behavior of $B_{c2,b}(T)$, including preliminary evidence for an $S$-shaped curve, strongly suggests something akin to a reentrant superconducting phase overlapping the low-field phase, which would be similar to the two phases of URhGe.  Fitting such behavior  will require significant modifications to the theory,  such as by including ferromagnetic fluctuations\cite{HattoriT}, field-dependent interactions\cite{HattoriK}, different FS shapes,\cite{Youngner,Rieck}, and two ferromagnetically-split FSs, which modifications are currently under study\cite{Lorscher2}.
Although an axial $p$-wave topological superconductor is presently elusive,  this theory could be useful to identify a future candidate material.

In summary, we have studied the two most-common versions of an axially-symmetric $p$-wave pair state, the Anderson-Brinkman-Morel (ABM) and Scharnberg-Klemm (SK) states.  For all induction ${\bm B}$ directions and temperatures $T$, the reduced (dimensionless) the SK state $B_{c2}(\theta,\phi,t)$  exceeds that of the ABM state.  Surprisingly, for $0\le\theta\le\theta^{*}$, the only $\theta$-dependence of $B_{c2}(\theta,\phi,t)$ arises from effective mass anisotropy, but then $B_{c2}(\theta)$  exhibits a kink  at $\theta^{*}[t,\gamma^2(\phi)]$.  Hence, it appears that there are two basic states evident in $b_{c2}(\theta,\phi,t)$: the nodal, chiral SK state for $-\theta^{*}\le\theta\le\theta^{*}$, and the antinodal, non-chiral polar state for $\theta^{*}\ge\theta\ge-\theta^{*}$.

\begin{acknowledgments}
The authors thank  J.-P. Brison, A. DeVisser, A. D. Huxley, Y. Matsuda, and K. Scharnberg for useful discussions.  This work was supported in part by
 the Florida Education Fund, the McKnight Doctoral Fellowship, a Chateaubriand Fellowship from the Embassy of France, UCF startup funds, the Specialized Research Fund for the Doctoral Program of Higher Education of China (no. 20100006110021) and by Grant no. 11274039 from the National Natural Science Foundation of China.

\end{acknowledgments}


\end{document}